\documentclass[%
reprint,
superscriptaddress,
showpacs,
preprintnumbers,
 bibnotes,
 amsmath,amssymb,
 aps,
prb,
]{revtex4-1}

\usepackage{graphicx}
\usepackage{dcolumn}
\usepackage{bm}
\usepackage{color}

\begin{document}


\title{Identification and mechanical control of ferroelastic domain structure in rhombohedral CaMn$_7$O$_{12}$}

\author{Renliang~Yuan}
\affiliation{International Center for Quantum Materials, School of Physics, Peking University, Beijing 100871, China}
\author{Lian~Duan}
\affiliation{International Center for Quantum Materials, School of Physics, Peking University, Beijing 100871, China}
\author{Xinyu~Du}
\affiliation{International Center for Quantum Materials, School of Physics, Peking University, Beijing 100871, China}
\author{Yuan~Li}
\email[]{yuan.li@pku.edu.cn}
\affiliation{International Center for Quantum Materials, School of Physics, Peking University, Beijing 100871, China}
\affiliation{Collaborative Innovation Center of Quantum Matter, Beijing 100871, China}


\begin{abstract}
We report on observation of ferroelastic domain structure in single crystals of multiferroic CaMn$_7$O$_{12}$ at room temperature. Two types of ferroelastic domain wall are found, consistent with the material's rhombohedral symmetry that is reduced from cubic symmetry at higher temperatures. Using Raman spectroscopy along with other measurements, we develop a systematic method to determine the microscopic domain orientation. Moreover, we find a switching behavior of the domains, which allows us to detwin the crystals conveniently at room temperature using a moderate uniaxial compression. Our result paves the way for further spectroscopic study and domain engineering in CaMn$_7$O$_{12}$.
\end{abstract}


\maketitle


\section{\label{sec:intro}Introduction}

Ferroic (ferroelectric, ferromagnetic, ferroelastic) domain walls have aroused persistent research interest due to their significance in both fundamental research and promising applications.\cite{catalan2012domain,salje2012ferroelastic,yang2014multifunctionalities} In so-called multiferroic materials, different ferroic order parameters coexist and exhibit mutual coupling, hence allowing for the manipulation of one ferroic property through another. This coexistence gives rise to composite domain walls \cite{fiebig2002observation,tokunaga2009composite,choi2010insulating} that might be a key to utilizing the mutual controllability of ferroic properties in applications. Furthermore, domain walls can exhibit distinctly different properties from the bulk,\cite{aird1998sheet,seidel2009conduction,he2012magnetotransport} leading to the possibility of using the domain walls as device. In addition to the intensively studied system BiFeO$_3$,\cite{catalan2009physics} intriguing domain structures have been observed and tuned in manganites.\cite{fiebig2002observation,meier2012anisotropic,choi2010insulating,jungk2010electrostatic,chae2012direct}

CaMn$_7$O$_{12}$ is a ``type-II'' multiferroic material \cite{KhomskiiPhysics2009} with very large ferroelectric polarization induced by magnetic order, and has been the subject of considerable recent research efforts.\cite{johnson2012giant,perks2012magneto,lu2012giant,iliev2014raman,du2014soft} Unlike in many widely studied ferroelectric materials, the occurrence of ferroelectricity in CaMn$_7$O$_{12}$ is preceded by a ferroelastic structural phase transition at higher temperatures, making it an ideal platform to study the ferroelastic and ferroelectric domain structures separately. At high temperatures, CaMn$_7$O$_{12}$ possesses the AC$_3$B$_4$O$_{12}$ cubic structure which is a derivative of simple perovskite ABO$_3$.\cite{vasil2007new} Upon cooling, a first-order phase transition occurs at $T_\mathrm{s} \approx$ 440 K. The four Mn$^{3.25+}$ ions in each formula unit are charge-ordered into three Mn$^{3+}$ and one Mn$^{4+}$ ions. The body diagonal of the high-temperature cubic cell that runs through the Mn$^{4+}$ ions shrinks a little bit, and becomes the $c$-axis of the new rhombohedral unit cell in hexagonal basis.\cite{bochu1980bond,przenioslo2002phase} This hexagonal $c$-axis plays an important role that not only determines the direction of the incommensurate orbital order established below $T_\mathrm{o} = 250$ K,\cite{perks2012magneto} but also sets the direction of the giant improper ferroelectric polarization that arises from the helical magnetic order below $T_{N1}$ = 90 K.\cite{johnson2012giant,lu2012giant} The pseudo-cubic cell below $T_\mathrm{s}$, along with the shortened body diagonal, are denoted by yellow/blue cubes and magenta lines, respectively, in Figs.~\ref{fig1}(a-b), omitting most details.

Since current methods of growing single crystals of CaMn$_7$O$_{12}$ all take place at temperatures far above the ferroelastic structural transition temperature $T_\mathrm{s}$,\cite{johnson2012giant,iliev2014raman,du2014soft} one can reasonably expect ferroelastic domains to form upon cooling the crystals to room temperature, as a result of simultaneous nucleation from different parts in the crystal, just like those in YBa$_2$Cu$_3$O$_{6+\delta}$ \cite{schmid1988polarized} and BaFe$_2$As$_2$.\cite{tanatar2009direct} In this paper, we focus on domain structures in CaMn$_7$O$_{12}$ at room temperature, where the compound is rhombohedral but not ferroelectric. We present direct observations of two types of stripe-like ferroelastic domain structures in single crystals. By using polarized-light microscopy, Raman spectroscopy and stylus surface profiler, we study the ferroelastic domain structures we observe. We propose a method to uniquely determine surface domain structures using Raman spectroscopy in combination with stylus profiler. Moreover, we identify a switching behavior of the ferroelastic domains, and show that the crystals can be mechanically detwinned by a moderate uniaxial compression. These findings are important for further spectroscopic studies of CaMn$_7$O$_{12}$ that require single-domain samples, and may facilitate future investigations of domain and domain-wall properties in multifunctional oxides.

\section{EXPERIMENTAL METHODS}

High-quality cube-shaped single crystals of CaMn$_7$O$_{12}$ were grown with a flux-reaction method \cite{johnson2012giant} at a cooling rate of 5 $^\circ$C/h. Natural facets of the crystals are parallel to crystallographic $\left\{100\right\}_\mathrm{c}$ planes, where the subscript ``c'' denotes pseudo-cubic notation. The samples were characterized as described elsewhere.\cite{du2014soft} Optical images of crystal surfaces were taken with a polarized-light microscope Olympus BX51 with polarizer and analyzer set in an almost perpendicular configuration. Differential interference contrast apparatus was installed to enhance image contrast. Surface profiles of crystal facets were measured with a KLA-Tencor P-6 Stylus Profiler using a contact force of 0.5 mg. Raman scattering measurements were performed in a back-scattering confocal geometry using the 632.8 nm line of a He-Ne laser for excitation. The diameter of the focused laser spot is estimated to be less than 5 microns. A Horiba Jobin Yvon LabRAM HR Evolution spectrometer, equipped with a 600 gr/mm grating and a liquid-nitrogen-cooled CCD detector, was used to analyze the Raman spectra.

\section{RESULTS AND DISCUSSIONS}

\subsection{Possible domain-wall orientations}\label{sec1}

\begin{figure}[htbp!]
\includegraphics{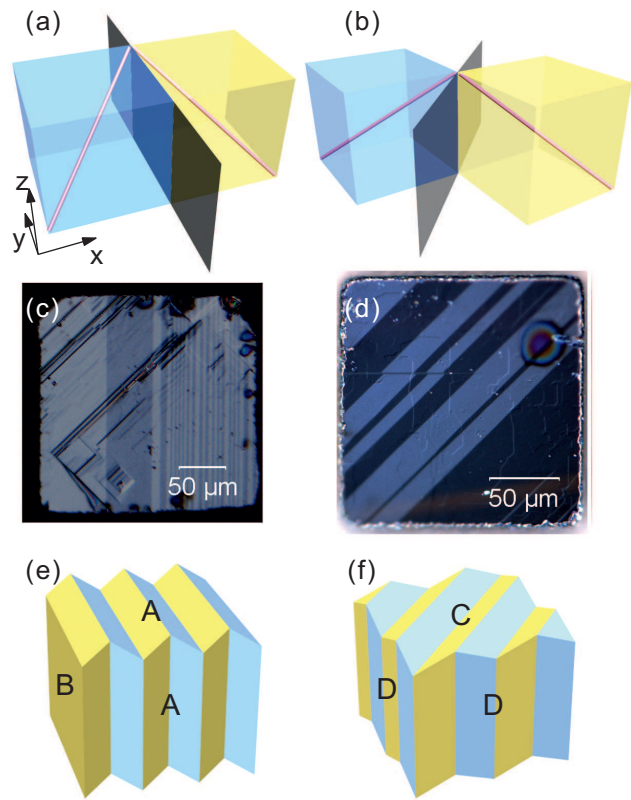}
\caption{\label{fig1}(Color online) (a), (b) Schematics of $\{100\}_\mathrm{c}$ and $\{110\}_\mathrm{c}$ type domain walls in CaMn$_7$O$_{12}$. The blue and yellow cubes indicate different domains with the magenta lines indicating the shortened body diagonals. Domain walls are denoted by the planes separating the cubes. (c), (d) Polarized-light optical images of single crystals with $\{100\}_\mathrm{c}$ and $\{110\}_\mathrm{c}$ domain walls. In addition to the domain structures which appear as regular bright-and-dark stripe patterns, growth terraces are seen especially in (c), but they do not seem to affect the domain distribution. (e), (f) Three-dimensional illustrations of the crystals in (c) and (d), respectively, with blue and yellow slabs indicating the two different domains. Surfaces with four inequivalent domain structures are labeled as A, B, C, and D.}
\end{figure}

Figures \ref{fig1}(c-d) display typical polarized-light optical images of multi-domain samples. Ferroelastic domains manifest themselves in a regular bright-and-dark stripe pattern. While only the crystals' top faces are shown, the domain structures actually extend to span the entire crystals, \textit{i.e.}, a consistent pattern is found on the side faces as well. This is illustrated (with exaggerated rhombohedral distortion of the domains) in Fig.~\ref{fig1}(e) and Fig.~\ref{fig1}(f), in which the slab-like domains are found to stack along the pseudo-cubic $\langle100\rangle_\mathrm{c}$ and $\langle110\rangle_\mathrm{c}$ directions, respectively. On slightly twinned samples, especially those grown at a cooling rate slower than 5 $^\circ$C/h, we can occasionally find stripes near the edges of the faces that do not span the entire crystal.

When the crystals are heated to temperatures above $T_\mathrm{s}$, the stripe patterns gradually disappear, consistent with the recovery of a single-domain cubic structure above $T_\mathrm{s}$. Cycling the temperature through $T_\mathrm{s}$ can each time lead to a completely different stripe pattern in rhombohedral phase, which indicates that the domain formation is not pinned by disorder or defects, and it in turn confirms the high quality of our samples. On the other hand, when the samples are cooled down to cryogenic temperatures, the stripe pattern remains even below $T_{N1}$ = 90 K in the ferroelectric phase. Hence it is possible to control low-temperature ferroelectric domain structures by pre-setting a desired ferroelastic domain structure in the paraelectric phase.

Grown at temperatures far above $T_\mathrm{s}$ and then cooled to room temperature, most crystals contain multiple ferroelastic domains due to simultaneous rhombohedral distortions that nucleates from different parts of the samples. Even though as-grown crystals can occasionally be found in a single-domain state,\cite{johnson2012giant,perks2012magneto} further manipulations may affect the domain structure. For example, to acquire pure $A_g$ Raman spectra in CaMn$_7$O$_{12}$, the preparation of a polished surface is required,\cite{du2014soft} and the polishing process will inevitably exert mechanical stress onto the sample, thus raising the risk of twinning it. We will discuss domain-switching behavior under external forces and how to utilize it to detwin crystals later.

Domain-wall orientations in ferroelastic materials can be understood by the equilibrium boundary condition (strain compatibility), which can be written as the following:\cite{sapriel1975domain}
\begin{equation}\label{eq1}
  \sum_{i,j=1}^3 [S_{ij}-S_{ij}']x_ix_j = 0,
\end{equation}
where $S$ and $S^\prime$ are the strain tensors of two adjacent domains. Indices $i$ and $j=$ 1, 2, and 3 denote Cartesian coordinates, and all possible $(x_1, x_2, x_3)$ that satisfy the equation constitute permissible boundaries between the two domains.

The rhombohedral phase of CaMn$_7$O$_{12}$ belongs to the ferroelastic species m3F$\overline{3}$ with four possible domain variants,\cite{aizu1969possible} which is the result of symmetry lowering from Im$\overline{3}$ to R$\overline{3}$.\cite{bochu1980bond} The four different spontaneous strain tensors can be written in the form:\cite{aizu1970determination}
\begin{equation}\label{eq2}
\begin{split}
  S_1 &= \left( \begin{array}{ccc} 0&d&d\\d&0&d\\d&d&0\\ \end{array} \right), \,\,\,\,\,\,\,\, S_2 = \left( \begin{array}{ccc} 0&-d&-d\\-d&0&d\\-d&d&0\\ \end{array} \right), \\
  S_3 &= \left( \begin{array}{ccc} 0&-d&d\\-d&0&-d\\d&-d&0\\ \end{array} \right), S_4 = \left( \begin{array}{ccc} 0&d&-d\\d&0&-d\\-d&-d&0\\ \end{array} \right),
\end{split}
\end{equation}
with the corresponding shortened body diagonals lying in direction $[111]_\mathrm{c}$, $[\overline{1}11]_\mathrm{c}$, $[1\overline{1}1]_\mathrm{c}$, and $[11\overline{1}]_\mathrm{c}$, respectively. When any two of these four domain variants meet, the solutions to Eq.~\ref{eq1} correspond to one of the equivalent crystallographic planes $\{100\}_\mathrm{c}$ and $\{110\}_\mathrm{c}$,\cite{sapriel1975domain} which is in perfect agreement with our observations in Fig.~\ref{fig1}. The two different types of domain walls are illustrated in Figs.~\ref{fig1}(a-b). Similar domain structures have been found in BiFeO$_3$,\cite{zavaliche2006multiferroic} BaTiO$_3$,\cite{marton2010domain} LaAlO$_3$,\cite{bueble1998influence} etc. All of these perovskites have rhombohedrally distorted phase, the strain tensors of which are in the same form \cite{aizu1970determination} as in our case. Additional domain variants may exist when ferroelectricity sets in, but the number of ferroelastic variants will remain to be four since 180$^\circ$ ferroelectric domain walls are not ferroelastic.\cite{sapriel1975domain,fousek1969orientation}

There are a total of four inequivalent surface domain structures on any $\{100\}_\mathrm{c}$ face of a cube-shaped crystal. They are labeled as A, B, C, and D in the illustrations in Figs.~\ref{fig1}(e-f). On an A-type face one finds $\langle100\rangle_\mathrm{c}$ stripes in the optical image, which continue onto adjacent faces as $\langle100\rangle_\mathrm{c}$ stripes; a B-type face exhibits no stripes since it is single-domain; a C-type face has $\langle110\rangle_\mathrm{c}$ stripes; a D-type face has $\langle100\rangle_\mathrm{c}$ stripes, but unlike the A-type, the stripes continue onto adjacent faces as $\langle110\rangle_\mathrm{c}$ stripes. Therefore, for a given crystal with sufficiently large single-domain volumes that span the entire crystal, one can obtain a rough idea about its domain structure by visually inspecting all of its six faces under a polarized-light microscope.

\subsection{Identification of individual domain orientation}\label{sec2}

\begin{figure}[!htbp]
\includegraphics{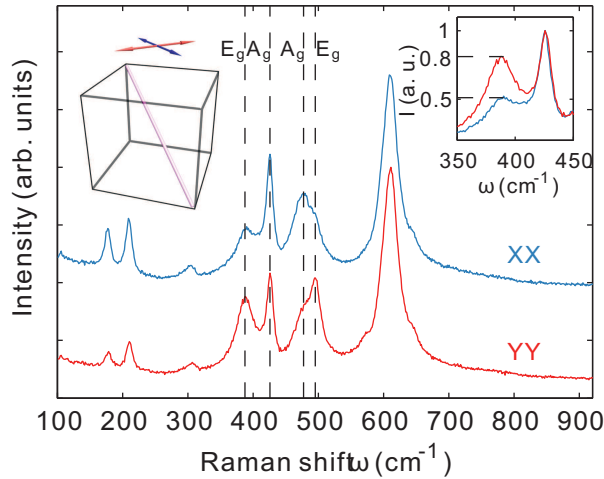}
\caption{\label{fig2}(Color online) Polarized Raman spectra obtained at room temperature, offset for clarity. Data curves are color-coded with arrows in the upper-left inset that indicate the polarization geometries with respect to the hexagonal $c$-axis (magenta body diagonal of the cube). Vertical dashed lines indicate $A_g$ and $E_g$ phonon peaks.\cite{du2014soft} Upper-right inset, XX and YY spectra in the 350 to 450 cm$^{-1}$ region, normalized at 425 cm$^{-1}$.}
\end{figure}

Optical images of crystal faces can only provide information about the \textit{relative} orientations of the hexagonal $c$-axis in adjacent domains (Figs.~\ref{fig1}(a-b)). Other methods are required to reveal the \textit{absolute} orientation of individual domains. We reported in our previous work\cite{du2014soft} that Raman scattering can detect $A_g$ and $E_g$ optical phonons separately with parallel and perpendicular combinations of incoming- and scattered-photon polarizations, respectively, when at least one of the polarizations is parallel to the hexagonal $c$-axis. If none of the polarizations is along the hexagonal $c$-axis, the acquired Raman spectrum will be a weighted combination of $A_g$ and $E_g$ signals.\cite{iliev2014raman} The relative intensities of $A_g$ and $E_g$ phonon peaks depend on the exact scattering geometry. Figure \ref{fig2} displays Raman spectra taken under ``XX'' and ``YY'' geometries on a single-domain crystal. Here, XX denotes that both the incoming- and scattered-photon polarizations are along the $[110]_\mathrm{c}$ direction, whereas YY denotes that both polarizations are along the $[1\overline{1}0]_\mathrm{c}$ direction; the hexagonal $c$-axis of the crystal is determined to be along the $[111]_\mathrm{c}$ direction (see below). The measurement configurations are illustrated in upper-left inset of Fig.~\ref{fig2}. Comparing the XX and YY spectra, one sees a clear difference in the 380 to 500 cm$^{-1}$ region, where four peaks can be attributed to $A_g$ and $E_g$ modes.\cite{du2014soft} The difference is due to the fact that both the incoming- and scattered-photon polarizations have a non-zero projection along the hexagonal $c$-axis in the XX scattering geometry. This non-zero projection leads to a larger (smaller) weight of the $A_g$ ($E_g$) signals in the Raman spectrum compared to that obtained in the YY scattering geometry, in which the photon polarizations are perpendicular to the hexagonal $c$-axis.

By performing both XX and YY measurements on only one face of the cube and after obtaining the data shown in Fig.~\ref{fig2}, we can only tell that the hexagonal $c$-axis is along either the $[111]_\mathrm{c}$ or the $[11\overline{1}]_\mathrm{c}$ direction. In order to unambiguously determine the hexagonal $c$-axis, one needs to perform additional measurements on adjacent faces of the cube. To verify the validity of our method, we have performed six pairs of XX and YY measurements on all faces of several single-domain crystals, all yielding consistent results supporting the aforementioned picture. This gives us confidence on how to identify and orient single-domain crystals. Moreover, it suggests the possibility to determine the orientations of individual domains in multi-domain samples using Raman spectroscopy.

\begin{figure*}[!htbp]
\includegraphics{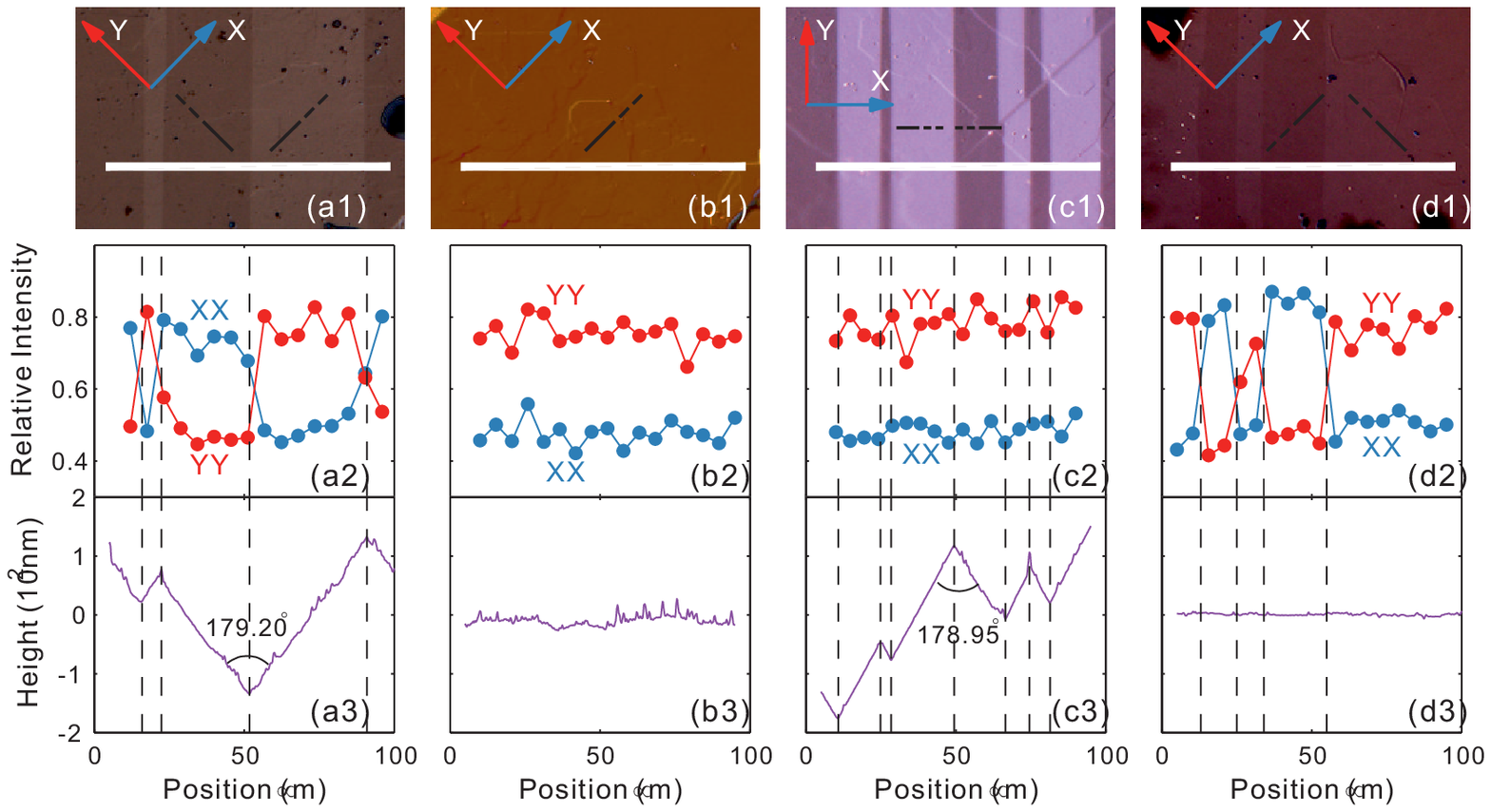}
\caption{\label{fig3}(Color online) (a1)-(d1) Optical images of A-, B-, C-, and D-type crystal faces (Figs. \ref{fig1}(e-f)). Horizontal white line indicates the trajectory along which the measurements in (a2)-(d2) and (a3)-(d3) are performed. Half-solid-half-dashed line indicates the orientation of the hexagonal $c$-axis (along one of the $\langle111\rangle_\mathrm{c}$ directions) in each domain, with the solid end pointing at the top face. (a2)-(d2) Raman intensity ratio $R$ (see text) measured along the trajectories indicated in (a1)-(d1), respectively, with different photon polarizations. The data are color-coded with the arrows indicating polarization directions in (a1)-(d1). (a3)-(d3) Surface profiles along the trajectories indicated in (a1)-(d1).}
\end{figure*}

For simplicity, here we use the relative intensity ratio between 390 and 425 cm$^{-1}$, $R = I(390\,\mathrm{cm}^{-1})/I(425\,\mathrm{cm}^{-1})$, to represent the key characteristics of the XX and YY Raman spectra in Figs.~\ref{fig2}. As can be seen from the upper-right inset, $R$ captures the most significant difference between the two types of spectra: it is approximately 0.8 or 0.5, respectively, when the photon polarizations are perpendicular to or partially along the hexagonal $c$-axis. To demonstrate how $R$ can be used to characterize surface-domain structures, we have performed space-resolved Raman scattering measurements on each of the four types of crystal faces that are labeled as A, B, C, and D in Figs.~\ref{fig1}(e-f). The data are displayed in Figs.~\ref{fig3}(a2-d2), along with polarized-light optical images of the surfaces in the top panels (a1-d1). In the optical images a horizontal white line indicates the trajectory on which the Raman spectra were taken. In these Raman measurements, the incident and scattered photon polarizations are always kept to be parallel, along one of the two perpendicular face diagonals ($[110]_\mathrm{c}$ and $[1\overline{1}0]_\mathrm{c}$, labeled as ``XX'' and ``YY'', respectively) on the surface.

Indeed, on A-type (Figs.~\ref{fig3}(a1-a2)) and D-type (Figs.~\ref{fig3}(d1-d2)) faces, $R$ is found to switch between 0.5 and 0.8 every time the scanning position crosses a boundary between the stripes, and whenever $R_\mathrm{XX}$ is around 0.5, $R_\mathrm{YY}$ is around 0.8 (and vise versa). The results are in perfect agreement with our expectations based on the model illustrated in Fig.~\ref{fig1}. On the other hand, $R$ is found to remain roughly constant in a given scattering geometry across the entire B-type (Figs.~\ref{fig3}(b1-b2)) and C-type (Figs.~\ref{fig3}(c1-c2)) faces, for different reasons: the B-type surfaces are single-domain, whereas on the C-type surfaces the hexagonal $c$-axes in different domains have the same projection onto the surface plane. These results render Raman spectroscopy, when used alone, unable to distinguish between A-type and D-type, and between B-type and C-type surfaces. In combination with polarized-light microscopy one can easily tell them apart, but neither optical inspection nor Raman spectroscopy is able to distinguish, \textit{e.g.}, between hexagonal $c$-axis orientations of $[111]_\mathrm{c}$ and $[11\overline{1}]_\mathrm{c}$, where the degeneracy is due to the fact that light is propagating along the $[001]_\mathrm{c}$ direction.

A feasible way to completely determine the domain structure by looking at \textit{only one} surface is to measure alternating inclinations, or wrinkles, on the surface. Figures \ref{fig3}(a3-d3) display surface profiles measured roughly along the same trajectory on which we took the Raman spectra. We find that the A-type and C-type surfaces exhibit clear zigzag profiles, whereas the B-type and D-type surfaces are essentially flat. Moreover, the angles of the zigzag profiles on the A-type ($\approx0.80(6)^\circ$) and C-type ($\approx1.05(6)^\circ$) surfaces are slightly different. To understand these results, we refer to the schematics in Fig.~\ref{fig1}. For crystals with $\{100\}_\mathrm{c}$ domain walls (Figs.~\ref{fig1}(a, e)), the strain tensors of the two domains are $S_1$ and $S_2$ (Eq.~\ref{eq2}); when they are contracted with the vector $(1,0,0)$, which lies within all A-type surfaces and crosses the domains, the outcomes of $S_1$ and $S_2$ ($(0,d,d)$ and $(0,-d,-d)$, respectively) have opposite projections along both the $[010]_\mathrm{c}$ and $[001]_\mathrm{c}$ directions. These are the normal directions of the A-type surfaces, and thus the surfaces are wrinkled. A similar argument can be used to explain the wrinkles on C-type surfaces (Figs.~\ref{fig1}(b, f)), where the strain tensors can be taken as $S_2$ and $S_3$, which are to be contracted with the vector $(1,-1,0)$ and then projected along $[001]_\mathrm{c}$. For D-type surfaces, we should instead use $(1,0,0)$ (or $(0,1,0)$) as the vector to be contracted with $S_2$ and $S_3$, but here the outcomes have the same projection along the $[010]_\mathrm{c}$ (or $[100]_\mathrm{c}$) direction, hence the D-type surfaces are not wrinkled despite the presence of domain walls. No wrinkles are expected on single-domain B-type surfaces. Using the room-temperature lattice constants of CaMn$_7$O$_{12}$,\cite{przenioslo2004charge} the angles of wrinkles on A- and C-type surfaces are calculated to be 0.75$^\circ$ and 1.05$^\circ$, respectively, which are in good agreement with our surface-profile data.

Taking the above results altogether, we suggest a new method to determine ferroelastic domain structures based on measurements of only one pseudo-cubic sample surface. The most reliable way is to use Raman spectroscopy combined with surface profile measurements, with polarized-light microscopy being a complementary but not necessary method. First, one needs to find the orientation of the domain walls by scanning in different directions and rotating photon polarizations in the Raman scattering measurement, aiming to maximize the contrast in both the Raman and surface profile data. Second, the surface domain structure can be know by comparing the data with the results shown in Fig.~\ref{fig3}. Third, from the surface profile data one can further tell apart the aforementioned $[111]_\mathrm{c}$ and $[11\overline{1}]_\mathrm{c}$ degenerated situations on A- and C-type surfaces, using the fact that shortened hexagonal $c$-axis always connects the valleys of the wrinkles. Our method is particularly useful when the sample is in thin-film form, or when the edges of crystals are not along a high-symmetry direction. While piezoresponse force microscopy \cite{balke2009electromechanical} is most commonly used to study ferroelectric domain structures, our method provides a route to monitoring the ferroelastic domains both above and below the ferroelectric transition temperature, which may help improve our understanding of the interplay between different ferroic order parameters in multiferroics. To detect nano-scale domains, tip-enhanced Raman spectroscopy \cite{atkin2012nano,domke2010studying} can be used to enhance the spatial resolution of our method.

\subsection{Domain switching and detwinning effect}\label{sec3}

\begin{figure}[!htbp]
\includegraphics[width=2.875in]{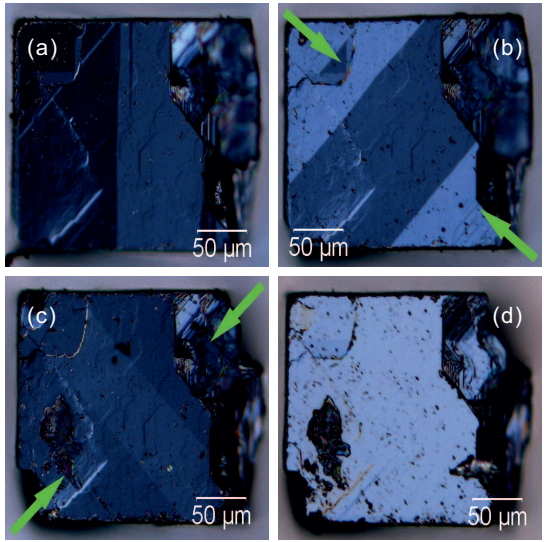}
\caption{\label{fig4}(Color online) Optical images of the same face of a crystal: (a) initial state, (b) after compression was applied along the face diagonal indicated by the arrows, (c) after compression was applied along the other face diagonal, and (d) detwinned state after compressed primarily along a body diagonal.}
\end{figure}

In addition to methods for characterizing the domains, we found that domain structure in CaMn$_7$O$_{12}$ single crystals can be altered at room temperature under moderate uniaxial compression. Figure~\ref{fig4} displays the same face of a crystal at different times. Initially, the surface exhibits a stripe pattern indicative of presence of $\{100\}_\mathrm{c}$ domain walls (Fig.~\ref{fig4}(a)). When a compressive force of about 1.5 N is applied along the direction shown by the arrows in Fig.~\ref{fig4}(b), the stripe pattern is rotated by 45$^\circ$, which indicates the formation of new $\{110\}_\mathrm{c}$ domain walls and the disappearance of the old ones. The magnitude of the applied force amounts to a uniaxial stress of about 30 MPa inside the crystal, and the new stripe pattern persists after the force was removed. Similarly, when a compression is applied along the direction in Fig.~\ref{fig4}(c), the stripe pattern is found to rotate again by 90$^\circ$. This domain-switching behavior can be understood as the following: when a uniaxial compression is applied, the ferroelastic domains rearrange themselves to minimize the length along the direction of compression. In the case of compression in the $[110]_\mathrm{c}$ direction, domains with hexagonal $c$-axes along $[111]_\mathrm{c}$ and $[11\overline{1}]_\mathrm{c}$ are energetically favored. According to the analysis in Section \ref{sec1}, this may result in the formation of either $(001)_\mathrm{c}$ or $(110)_\mathrm{c}$ domain walls, but the former would also generate wrinkles on the $(100)_\mathrm{c}$ and $(010)_\mathrm{c}$ faces which are incompatible with the applied compression. Hence the resulting domain walls are parallel to $(110)_\mathrm{c}$. For the same reason, one can reasonably expect that a $\langle111\rangle_\mathrm{c}$ domain can be exclusively selected if the compression is primarily along a body diagonal of the cube, as is indeed shown to be the case in Fig.~\ref{fig4}(d). Apart from taking optical images, we have performed Raman scattering measurements on all six faces of the sample in Fig.~\ref{fig4}(d), which consistently show that the crystal is highly detwinned.

This switching behavior of ferroelastic domains in CaMn$_7$O$_{12}$ stems from the fact that the rhombohedral distortion is characterized by  the shortening rather the  elongation of a body diagonal, which makes the detwinning operations simple. Among other rhombohedrally distorted perovskites, LaAlO$_3$ \cite{fay1967reorientation} is similar to CaMn$_7$O$_{12}$, whereas the distortion in BiFeO$_3$ features an elongated body diagonal, making it rather tricky to prepare certain types of domain structure.\cite{kubel1993growth,chu2009nanoscale}

\section{CONCLUSIONS}

In summary, we have observed and investigated ferroelastic domain structures in single crystals of CaMn$_7$O$_{12}$. For cube-shaped single crystals with multiple domains, we can determine the orientation of individual domains by measurements either on two adjacent faces with Raman spectroscopy alone, or on only one face using both Raman spectroscopy and stylus surface profiler. The latter method is also suitable for determining domain structures in thin-film samples, where only one surface is available. In addition, polarized-light microscopy provides a complementary and convenient way to observe the domain structure. Finally, we find that the domain structure can be altered by moderate uniaxial compression at room temperature, which allows for a simple method to obtain twin-free samples with a controlled orientation of the hexagonal $c$-axis. Our results offer the opportunity to prepare well-defined CaMn$_7$O$_{12}$ samples, \textit{e.g.}, for spectroscopic studies that require single-domain crystals, and the methods we use can be readily transferred to studies of thin-film samples as well as other rhombohedrally distorted cubic compounds.

\begin{center}
$\textbf{Acknowledgments}$
\end{center}

We thank W. H. Yang and X. B. Wang for technical assistance in the use of polarized-light microscope and stylus profiler. This work is supported by the NSF of China (No. 11374024) and the NBRP of China (No. 2013CB921903).

\bibliography{cmoref}

\end{document}